\documentclass[aps,prl,reprint,superscriptaddress]{revtex4-1}
\usepackage{graphicx}
\usepackage{booktabs}
\usepackage{amssymb}
\usepackage{graphics}
\usepackage{amsmath}
\usepackage{placeins}
\usepackage{hyperref}
\usepackage{verbatim}

\begin{document}

%\preprint{}

\title{Experimental demonstration of the correction of coupled transverse dynamics aberration in an rf photoinjector}


\author{Lianmin Zheng}
%\thanks{}
\affiliation{Department of Engineering Physics, Tsinghua University, Beijing 100084, People's Republic of China}
\affiliation{Key Laboratory of Particle and Radiation Imaging, Tsinghua University, Ministry of Education, Beijing 100084, People’s Republic of China}
\affiliation{High Energy Physics Division, Argonne National Laboratory, Lemont, Illinois 60439, USA}
\author{Jiahang Shao}
\affiliation{High Energy Physics Division, Argonne National Laboratory, Lemont, Illinois 60439, USA}
\author{Yingchao Du}
\email[]{dych@mail.tsinghua.edu.cn}
\affiliation{Department of Engineering Physics, Tsinghua University, Beijing 100084, People's Republic of China}
\affiliation{Key Laboratory of Particle and Radiation Imaging, Tsinghua University, Ministry of Education, Beijing 100084, People’s Republic of China}
\author{John G. Power}
\author{Eric E. Wisniewski}
\author{Wanming Liu}
\author{Charles E. Whiteford}
\author{Manoel Conde}
\author{Scott Doran}
\affiliation{High Energy Physics Division, Argonne National Laboratory, Lemont, Illinois 60439, USA}
\author{Chunguang Jing}
\affiliation{High Energy Physics Division, Argonne National Laboratory, Lemont, Illinois 60439, USA}
\affiliation{Euclid Techlabs LLC, Bolingbrook, Illinois 60440, USA}
\author{Chuanxiang Tang}
\affiliation{Department of Engineering Physics, Tsinghua University, Beijing 100084, People's Republic of China}
\affiliation{Key Laboratory of Particle and Radiation Imaging, Tsinghua University, Ministry of Education, Beijing 100084, People’s Republic of China}
\author{Wei Gai}
\affiliation{Department of Engineering Physics, Tsinghua University, Beijing 100084, People's Republic of China}
\affiliation{Key Laboratory of Particle and Radiation Imaging, Tsinghua University, Ministry of Education, Beijing 100084, People’s Republic of China}

%\collaboration{Jiahang Shao, John G. Power,}


\date{\today}

\begin{abstract}
The production of electron bunches with low transverse emittance approaches the thermal emittance of the photocathode as various aberrations are corrected.  Recently, the coupled transverse dynamics aberration was theoretically identified as a significant source of emittance growth and a corrector magnet was proposed for its elimination~[D.H. Dowell, F. Zhou, and J. Schmerge, PRAB \textbf{21}, 010101 (2018)]. This aberration arises when the beam acquires an asymmetric distribution that is then rotated with respect to the transverse reference axis thus introducing a correlation in the vertical and horizontal planes.  The asymmetry is introduced by a weak quadrupole field in the rf gun or emittance compensation solenoid and the rotation is caused by the solenoid.  This Letter presents an experimental study of the coupled transverse dynamics aberration in an rf photoinjector and demonstrates its elimination by a quadrupole corrector consisting of a normal and a skew quadrupole.  The experimental results agree well with theoretical predictions and numerical simulations. The study also demonstrates the emittance of a low charge beam can be preserved during transportation at its thermal value, which was 1.05~mm\,mrad/mm, for the cesium telluride photocathode and 248~nm UV laser used.
\end{abstract}

%\keywords{}

\maketitle


\par Normalized transverse emittance describes the beam size in phase space and determines the beam brightness at a fixed charge. It is a key figure of merit for high-brightness accelerators based on electron injectors such as: X-ray free electron laser~\cite{emma2010first,ackermann2007operation}, electron-positron linear collider~\cite{agapov2007beam,accomando2004physics}, ultrafast electron diffraction and microscopy~\cite{weathersby2015mega,li2009experimental}, Thomson scattering X-ray source~\cite{du2013generation,gibson2010design}, etc.

\par The lowest achievable transverse emittance equals the thermal (or intrinsic) emittance of the photocathode, $\varepsilon _{therm}$. However, \textit{other} mechanisms such as space charge~\cite{limborg2006optimum}, rf field~\cite{chae2011emittance}, and spherical/chromatic aberrations~\cite{dowell2016sources,mcdonald1989frontiers} can lead to emittance growth.  Much of the research in high-brightness injectors over the last decades has focused on: (1) reducing the thermal emittance of the cathode~\cite{Karkare2017Reduction,N2010High} and (2) identifying sources of emittance growth and developing methods to eliminate them.  These methods include emittance compensation using a solenoid near the gun~\cite{Carlsten1989New}, rf-symmetrized gun design~\cite{zheng2016development,Xiao2005Dual}, and bunch profile optimization~\cite{Luiten2004How,Li2009Laser}. 

\par Recently, a new aberration was identified~\cite{dowell2018exact} as a source of emittance growth in high brightness injectors called the quadrupole-coupled transverse dynamics aberration, or coupled aberration for short. This coupled aberration occurs when the beam acquires a quadrupole distribution that is rotated (with respect to the transverse reference axis) by a solenoid that couples the motion between the $x-x'$ and the $y-y'$ planes.  Although the 4D emittance remains unchanged, the 2D emittance can be significantly increased due to the coupled aberration~\cite{dowell2018exact,zheng2018overestimation}.  The total 2D emittance is the quadrature sum of the thermal emittance and the emittance contributions due to the various aberrations, 
\begin{equation}\label{total_emittacne}
{\varepsilon} = \sqrt {{\varepsilon _{therm}}^2 + {\varepsilon _{coupled}}^2 + {\varepsilon _{other}}^2} 
\end{equation}
where $\varepsilon _{coupled}$ and $\varepsilon _{other}$ are the emittance growth from the quadrupole-coupled transverse dynamics aberration and the aforementioned \textit{other} emittance sources, respectively.

\par Previous theoretical studies have shown that a quadrupole corrector, which produces a rotated quadrupole field, can be used to reduce the emittance growth from the coupled aberration~\cite{dowell2018exact,zheng2018overestimation}. The reduction depends on the strength and the rotation angle of the applied quadrupole corrector field.  In addition to theoretical studies, several groups have developed quadrupole correctors for photoinjectors running with a bunch charge between  100~pC and 500~pC~\cite{dowell2018exact,bartnik2015operational,schietinger2016commissioning,krasilnikov2018electron}.  However, due to the relatively high charge,  $\varepsilon _{other}$ cannot be distinguished from $\varepsilon _{coupled}$. This limits the physical understanding of the quadrupole-coupled transverse dynamics aberrations as well as the validation of the corrector optimization and application.

\par  In this Letter, we present a systematic experimental study of the coupled aberration and its correction using a quadrupole corrector. During the experiment, the coupled aberration came from two sources: (1) a quadrupole rf field in the gun due to the asymmetry of the cavity followed by a rotation in the solenoid~\cite{dowell2018exact,zheng2018overestimation} and (2) a quadrupole dc field in the solenoid due to the asymmetry of the solenoid's yoke and/or coil~\cite{Xiao2005Dual,zheng2018overestimation,dowell2018exact}.  These two sources are referred to as the gun quadrupole and the solenoid quadrupole, respectively. 

\par In the presence of both the gun quadrupole and the solenoid quadrupole, the emittance growth due to the coupled aberrations can be expressed as~\cite{dowell2018exact,zheng2018overestimation}
\begin{equation}\label{emittacne_coupled}
\varepsilon _{coupled} = \beta \gamma \frac{{{\sigma _{x,sol}}{\sigma _{y,sol}}}}{{{f_a}}}\Big| {\sin 2(KL + \alpha_a )} \Big|
\end{equation}
where $f_a$ and $\alpha_a$ denote the combined strength and rotation angle of the quadrupole aberration field at the solenoid entrance calculated with transfer matrix,  $\sigma _{x,sol}$ and $\sigma _{y,sol}$ are the rms beam sizes in horizontal and vertical directions inside the solenoid using the thin-lens approximation, and $KL$ denotes the Larmor angle of the solenoid, respectively. 

\par If a quadrupole corrector with focal length $f_{c}$ and rotation angle $\alpha_{c}$ is placed near the solenoid to reduce the coupled emittance, the emittance growth after correction can be expressed as~\cite{dowell2018exact,zheng2018overestimation}
\begin{equation}
\label{emittance growth}
\begin{aligned}
\varepsilon _{coupled-correction} = &\beta \gamma \Big|
\displaystyle{\frac{{{\sigma _{x,sol}}{\sigma _{y,sol}}}}{{{f_a}}}}\sin 2( {KL+{\alpha _a}})\\
&+\frac{{{\sigma _{x,c}}{\sigma _{y,c}}}}{{{f_{c}}}}}\sin 2{\alpha _{c} \Big|
\end{aligned}
\end{equation}
where $\sigma _{x,c}$ and $\sigma _{y,c}$ are the rms beam sizes in horizontal and vertical direction inside the quadrupole corrector using thin-lens approximation. By choosing proper $f_{c}$ and $\alpha_{c}$, the aberration can be fully eliminated with $\varepsilon _{coupled-correction}$ reduced to zero.

\begin{figure}[hbtp]
	\centering
	\includegraphics[scale=0.55]{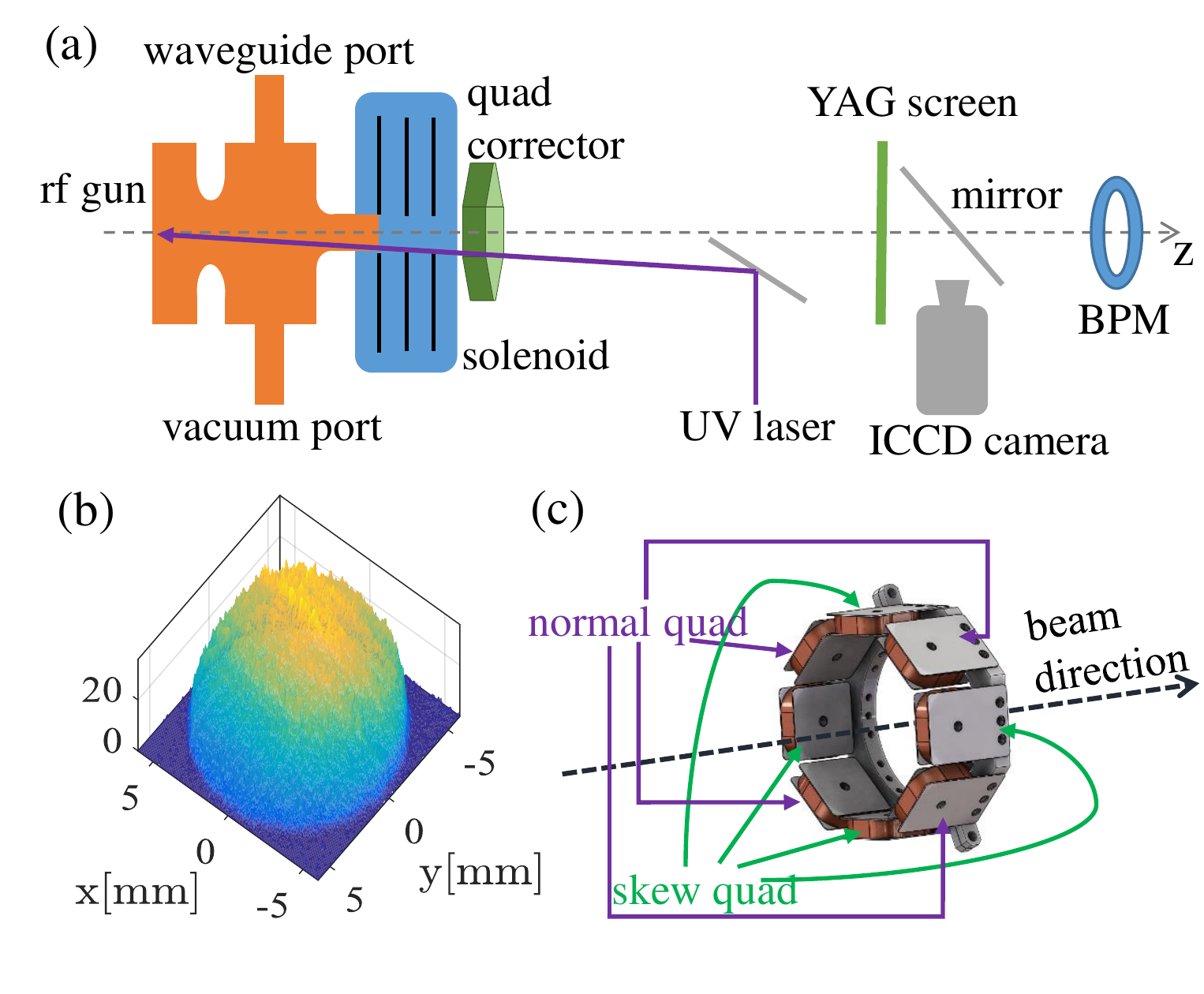}
	\caption{\label{Fig.setup} The experimental setup at AWA. (a) The layout of the beamline. (b) The intensity distribution of the laser spot homogenized by the microlens arrays. (c) The design of the quadrupole corrector.}
\end{figure}

\par The experimental study was conducted with the L-band 1.6-cell rf photocathode gun at the Argonne Wakefield Accelerator (AWA) facility, as shown in Fig.~\ref{Fig.setup}(a). A 248~nm UV laser was applied to generate electrons from the 
cesium telluride cathode. The laser transverse distribution was homogenized with a microlens array to be relatively uniform~\cite{halavanau2017spatial}, as shown in Fig.~\ref{Fig.setup}(b). The electric field on the cathode was set to 32~MV/m to reduce the dark current background to facilitate acquiring the low-charge bunch profile on the YAG screen. The electron beam was launched at $43^\circ$ rf phase and its energy at the gun exit was 3.2~MeV. A solenoid after the gun was used to focus the beam onto a YAG screen perpendicular to the beamline for the emittance measurement. A PI-MAX Intensified CCD (ICCD) camera~\cite{camera} was used to capture beam images on the retractable YAG screen with a shutter width of 100~ns to improve the signal-to-noise ratio.  The spatial resolution of the system was $\sim$60~$\mu$m. A calibrated strip line beam position monitor (BPM) downstream was used to measure the charge with a sensitivity of $\sim$40~mV/1~pC.

\par The solenoid scan technique, in which the measured emittance is determined by fitting the beam size as a function of the solenoid strength~\cite{qian2012experimental,hauri2010intrinsic,maxson2017direct}, was used in this research. The solenoid scan range used in this study was only $\sim5\%$ of the average solenoid strength so that the maximum beam size is only about twice the minimum beam size; this reduces the fitting error~\cite{houjunphdthesis}. Within the small scan range, the variation of the gun quadrupole and solenoid quadrupole can be neglected.

\par The beam parameters were optimized to minimize $\varepsilon _{other}$ in the experiment in order to highlight the coupled aberration. (i) The space charge effect was minimized during the experiment by using a low-charge beam. The charge was gradually reduced until the measured emittance did not change. The charge used in the study was $\sim$1~pC for the largest root-mean-square (rms) spot size of 2.7~mm and the charge density was kept the same for smaller sizes. The following parameters were minimized via ASTRA simulations~\cite{floettmann2011astra}. (ii) The emittance growth due to the phase-dependent transverse kick in the rf gun, which is proportional to the beam length~\cite{chae2011emittance}, was reduced by using a short beam (1.5~ps FWHM) generated by short laser pulse. Based on simulations, the resulted emittance growth is less than 1.4\%. (iii) The low charge and short pulse length yielded a narrow energy spread of 0.1\% which contributes only 1.6\% emittance growth due to the chromatic aberration according to simulations. (iv) The beam size inside the solenoid was kept under 4~mm, which is much smaller than the 40~mm solenoid bore, so no spherical aberration was observed in simulation. Overall, the total contribution to the measured emittance from other sources is less than 3\% and is neglected in the study.

\par
\begin{figure}[hbtp]
	\centering
	\includegraphics[scale=0.8]{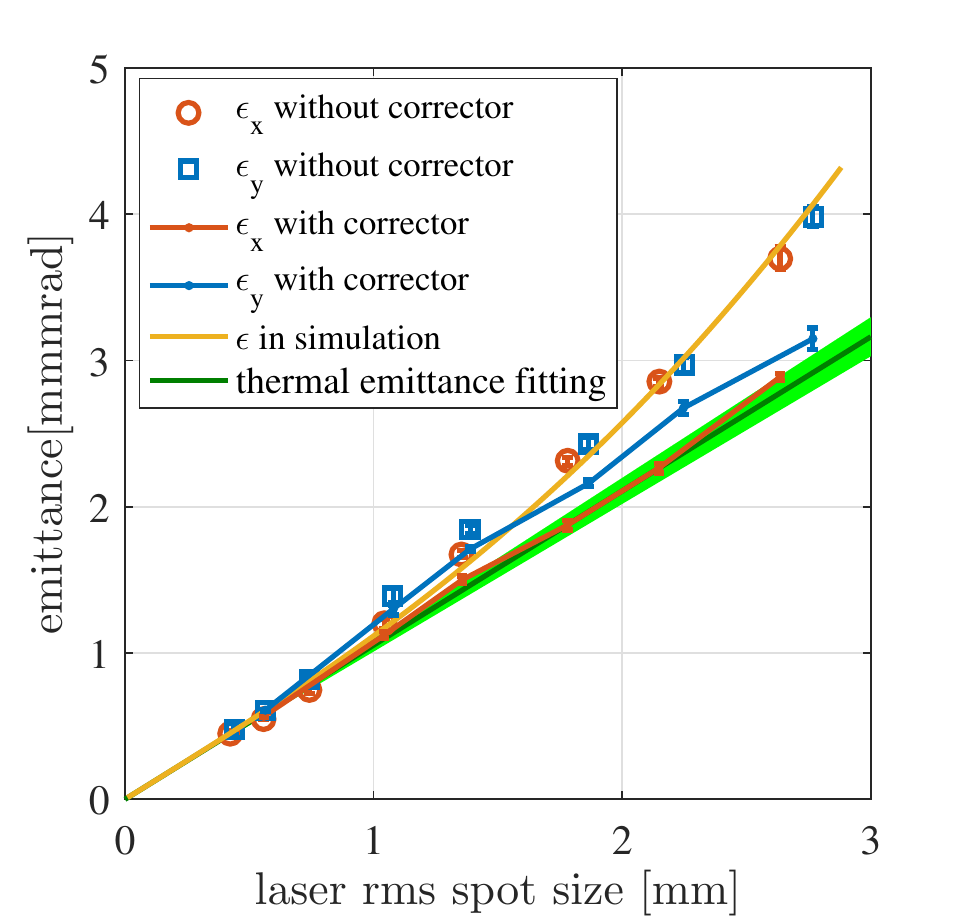}
	\caption{\label{Fig.emt_beamsize} Transverse emittance as a function of laser rms spot size. Circle and square dots: measured results without quadrupole corrector. Solid green line: thermal emittance value with error bar, extrapolated from the linear regime in measurement. Solid yellow line: simulated emittance by ASTRA without quadrupole corrector. Solid red and blue lines: measured results with quadrupole corrector set to proper strength and rotation angle to fully cancel the coupled transverse dynamics aberration.}
\end{figure}

 In the first part of the study, the emittance without the quadrupole corrector was measured for various laser spot sizes as shown in Fig.~\ref{Fig.emt_beamsize}. The results can be classified into two regimes: a linear regime where the rms spot size is small ($<$0.75~mm) and a nonlinear regime where the spot size is large ($>$0.75~mm). In the linear regime, the thermal emittance can be estimated to be $1.05\pm0.04$~mm\,mrad/mm from the slope of the line. This value is consistent with the theoretical estimate via the three-step emission model~\cite{flottmann1997note} by assuming the surface barrier to be 3.5~eV~\cite{dowell2010cathode,miltchev2005measurements} and field enhancement factor to be 1~\cite{zheng2018overestimation}. In the nonlinear regime, however, the measured emittance deviates from its thermal value as can be observed as the beam size increases.  This emittance growth is due to the coupled aberration as expected by Eqn~\ref{emittacne_coupled}. The largest emittance growth reaches $\sim$35\% with a laser rms spot size of 2.7~mm. The aberration was due to both the gun quadrupole and the solenoid quadrupole. The quadrupole mode of the rf gun has a relative strength of $7.2 \times 10^{-3}$ to the monopole mode due to the asymmetric waveguide/vacuum port design. The solenoid has a quadrupole component has a strength of 77~Gauss/m with 0.1974~T solenoid field and a rotation angle of 12~$^{\circ}$ as previously measured.~\cite{zheng2018overestimation}.

\par In the second part of the experiment, a quadrupole corrector was installed at the solenoid exit as illustrated in Fig.~\ref{Fig.setup}(a). The corrector consists of a normal and a skew quadrupole, made of eight coils, secured on an aluminum frame wrapped around the beam pipe (Fig.~\ref{Fig.setup}(c)). By independently controlling the strength of the two quadrupoles, the corrector's overall strength $k_c$ and rotation angle $\alpha_c$ can be varied as
\begin{equation}
\label{quad_strength n angle}
\left\{
\begin{aligned}
k_c=&\sqrt{k_{\rm normal}^{2}+k_{\rm skew}^{2}}\\
\alpha_c=&\frac{\arctan{(k_{\rm skew}/k_{\rm normal})}}{2}
\end{aligned}
\right.
\end{equation}
where $k_{\rm normal}$ and $k_{\rm skew}$ denotes the strength of the normal and the skew quadrupole, respectively. Here the corrector's strength can be expressed as $k_c\equiv \frac{{\partial {B_y}}}{{\partial x}}{\big|_{x,y = 0}}=\frac{{\beta \gamma mc}}{{{f_c}e{L_{\rm quad}}}}$, where $L_{\rm quad}$ is the effective length of the quadrupole corrector.

\par
\begin{figure*}[hbtp]
\centering
\includegraphics[scale=0.72]{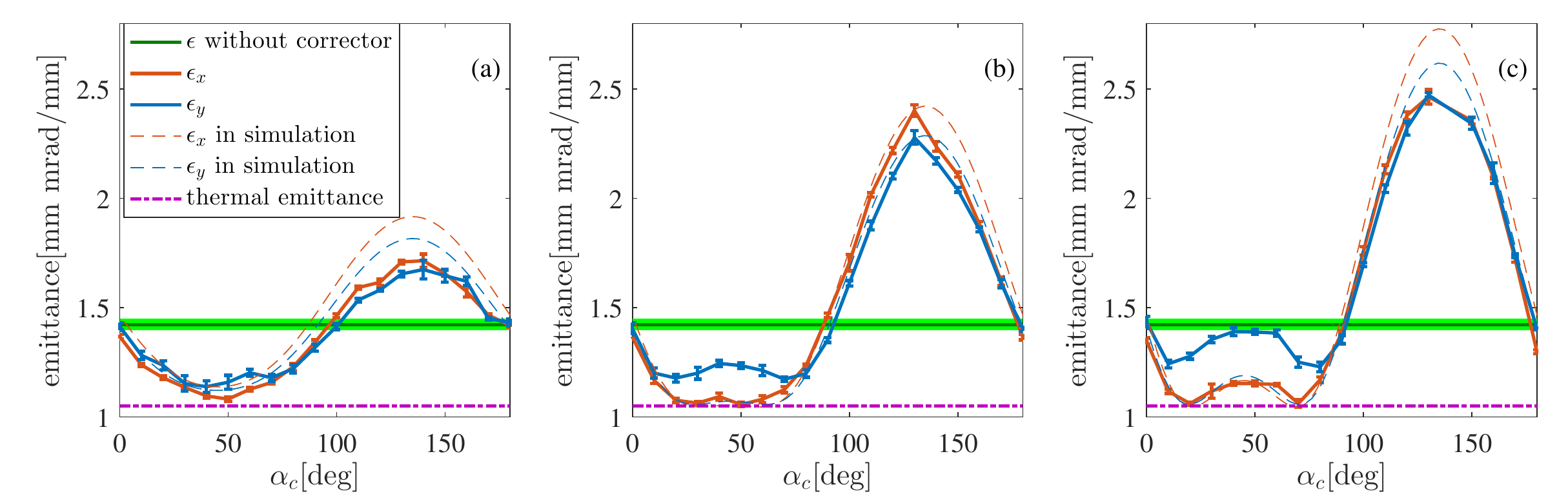}
\caption{\label{Fig.emt_theta} Measured transverse emittance for a 2.7~mm rms laser spot size as a function of the quadrupole corrector's strength and rotation angle. The green line represents the emittance without the corrector and its width denotes the measurement error bar. The dash magenta line represents the thermal value. The corrector strength in (a-c) is 36~Gauss/m, 72~Gauss/m, and 96~Gauss/m, respectively.}
\end{figure*}

 The strength and rotation angle of the quadrupole corrector (Eqn~\ref{quad_strength n angle}) was scanned in order to find the setting that would fully cancel the emittance growth (Fig.~\ref{Fig.emt_theta}). The measured emittance oscillates as a function of $\alpha_c$ and emittance reduction is observed when $0^{\circ}<\alpha_c<90^{\circ}$. In Fig.~\ref{Fig.emt_theta}(a), the corrector strength is weak (36 Gauss/m) and only one peak around $\alpha_c=135^{\circ}$ in the range of $0^{\circ}<\alpha_c<180^{\circ}$ is observed. On the other hand, in Fig.~\ref{Fig.emt_theta}(b) and (c), when the corrector strength is strong (72 and 96 Gauss/m), a second peak appears near  $\alpha_c=45^{\circ}$. In these cases, the minimum emittance in x-direction reaches the thermal value which demonstrates full cancellation of the emittance growth due to the coupled aberration. However, in the y-direction, although the emittance shows similar trend as the x-direction, its minimum value is larger. In addition, the amplitude of the deviation between the two directions increases with corrector strength.   Finally, the quadrupole corrector settings were scanned to fully cancel the aberration for various laser spot sizes. The minimum emittance after optimization is close to the thermal value for all spot sizes as illustrated in Fig.~\ref{Fig.emt_beamsize}.

\par The emittance measurement results discussed above were benchmarked with ASTRA simulations~\cite{floettmann2011astra}. The simulations included both the gun and solenoid qudrurpole as well as used a realistic laser transverse distribution (Fig.~\ref{Fig.setup}(b)).  The 3D field map of the photocathode rf gun, including the gun quadrupole component, was simulated with CST Microwave Studio~\cite{CST} and the solenoid quadrupole was modeled as having the same longitudinal profile as the solenoid while its strength and rotation angle were set to the values used during the experiment.

\par The ASTRA simulation results agree well with the emittance measurements made in both parts of the experiment: without and with the quadrupole corrector.  The simulated uncorrected emittance (without the corrector) as a function of laser spot size agrees well with the measurement, as illustrated by the yellow line in Fig.~\ref{Fig.emt_beamsize}. In addition, the simulated emittance as a function of the corrector strength and rotation angle in both the x- and y-directions (Fig.~\ref{Fig.emt_theta}, dashed red and blue lines) agrees well with the measured x-direction emittance but has some deviation from the measured y-direction emittance.  This may be caused by other mechanisms which are beyond the scope of this paper, such as aberrations from sextupole components~\cite{dowell2018correcting}.

\par
\begin{figure}[hbtp]
\centering
\includegraphics[scale=0.79]{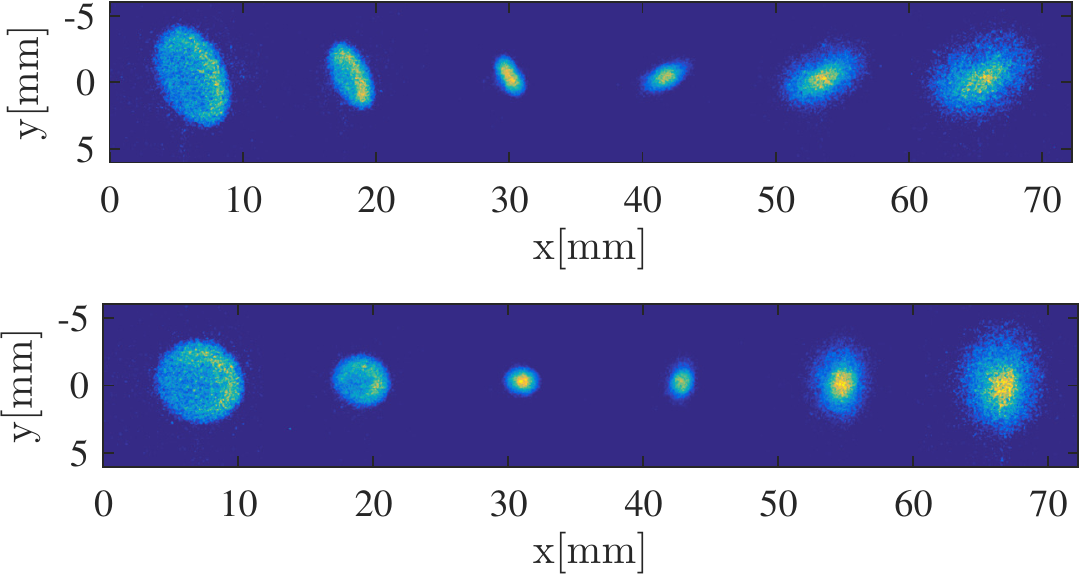}
\caption{\label{FIG.images} Transverse beam profile captured on the YAG screen during the measurement with the solenoid strength increasing from left to right (under-focus to over-focus). The rms laser spot size is 2.7~mm. Top: Without the quadrupole corrector. The measured emittance is 1.42~mm\,mrad/mm. Bottom: With the quadrupole corrector. The strength is 72~Gauss/m and the rotation angle is 70$^\circ$ and the measured emittance is 1.14~mm\,mrad/mm.}
\end{figure}

 In addition to the clear effect the quadrupole corrector had on the measured emittance (see above), it also had a noticeable effect on the transverse profile of the bunch during the solenoid scan, as illustrated in Fig.~\ref{FIG.images}. Without correction (top row) the bunch shape is tilted, which indicates an x-y coupling, as the solenoid strength is scanned. However, at optimized quadruple correction setting, the bunch shape becomes upright and the coupling is eliminated. It should be noted that the shape after correction is not necessarily round, since the corrector does not cancel the aberration locally, but the shape is not tilted.  This presents a simple way of finding the proper quadrupole corrector setting without performing the emittance measurements. 

\par In conclusion, a quadrupole corrector was installed on the L-band rf photoinjector at the AWA facility to study the coupled transverse dynamics aberration and to demonstrate elimination of the aberration. An electron bunch with low charge and short length was used to minimize the emittance growth from other sources. The measurement results are in good agreement with theoretical prediction and numerical simulation.  The results show that the coupled transverse dynamics aberration from several sources can be fully canceled with a single quadrupole corrector of the proper strength and rotation angle. In summary, this work studied a significant source of emittance growth and demonstrated an effective method for correcting the aberration. These results are expected to benefit the high-brightness, electron-injector-based R\&D accelerator community.

\begin{acknowledgments}
  \par  This work is supported by the U.S. Department of Energy, Offices of HEP and BES, under Contract No. DE-AC02-06CH11357. It is also funded by the National Natural Science Foundation of China (NSFC) under Grant No. 11435015 and No. 11375097.
\end{acknowledgments}

\bibliography{apstemplate.bib}

\end{document}